\documentclass[%
 aip,
 amsmath,amssymb,
 reprint,%
]{revtex4-1}

\usepackage{graphicx}
\usepackage{dcolumn}
\usepackage{bm}

\usepackage[utf8]{inputenc}
\usepackage[T1]{fontenc}
\usepackage{mathptmx}
\usepackage{etoolbox}
\usepackage{xcolor}

\makeatletter
\def\@email#1#2{%
 \endgroup
 \patchcmd{\titleblock@produce}
  {\frontmatter@RRAPformat}
  {\frontmatter@RRAPformat{\produce@RRAP{*#1\href{mailto:#2}{#2}}}\frontmatter@RRAPformat}
  {}{}
}%
\makeatother
\begin{document}

\preprint{AIP/123-QED}

\title[An Accessible Instrument for Measuring Soft Material Mechanical Properties]{An Accessible Instrument for Measuring Soft Material Mechanical Properties}
\author{B.M. Unikewicz}
 \affiliation{Massachusetts Institute of Technology, Department of Mechanical Engineering, \\ Cambridge, MA, 02139, USA}
 
\author{A.M. Pincot}%
\affiliation{Massachusetts Institute of Technology, Department of Mechanical Engineering, \\ Cambridge, MA, 02139, USA}
 
\author{T. Cohen}
\email{talco@mit.edu}
\affiliation{Massachusetts Institute of Technology, Department of Mechanical Engineering, \\ Cambridge, MA, 02139, USA}
\affiliation{Massachusetts Institute of Technology, Department of Civil \& Environmental Engineering, \\ Cambridge, MA, 02139, USA}

\date{\today}

\begin{abstract}
Soft material research has seen significant growth in recent years, with emerging applications in robotics, electronics, and healthcare diagnostics where understanding material mechanical response is crucial for precision design. Traditional methods for measuring nonlinear mechanical properties of soft materials require specially sized samples that are extracted from their natural environment to be mounted on the testing instrument. This has been shown to compromise data accuracy and precision in various soft and biological materials. To overcome this, the Volume Controlled Cavity Expansion (VCCE) method was developed. This technique tests soft materials by controlling the formation rate of a liquid cavity inside the materials at the tip of an injection needle, and simultaneously measuring the resisting pressure which describes the material response. Despite VCCE's early successes, expansion of its application beyond academia has been hindered by cost, size, and expertise. In response to this, the first portable, bench-top instrument utilizing VCCE is presented here. This device, built with affordable, readily available components and open-source software, streamlines VCCE experimentation without sacrificing performance or precision. It is especially suitable for space-limited settings and designed for use by non-experts, promoting widespread adoption. The instrument's efficacy was demonstrated through testing Polydimethylsiloxane (PDMS) samples of varying stiffness. This study not only validates instrument performance, but also sets the stage for further advancements and broader applications in soft material testing. All data, along with acquisition, control, and post-processing scripts, are made available on GitHub.
\end{abstract}

\maketitle

\section{\label{sec:level1}Introduction}
Soft materials have been an active area of research within academia and industry alike. Fields traditionally reliant on rigid materials, such as robotics \cite{kim2013soft, majidi2014soft, roche2014bioinspired} and electronics \cite{harris2016flexible, liu2017lab, rogers2010materials, li2023review, ziaie2004hard} are increasingly adopting soft materials due to their adaptability and utility in anthropomimetic design. In the field of biology, mechanics research has contributed to disease detection, \cite{yeh2002elastic, paszek2005tensional, samani2007inverse, last2011elastic, budday2015mechanical} food science, \cite{finney1967dynamic,solomon2007modeling} and tissue engineering, \cite{engler2004myotubes, kong2005non, vedadghavami2017manufacturing} which has opened new fields in organ 3D printing \cite{shopova2023bio, jammalamadaka2018recent, murphy20143d, radenkovic2016personalized} and understanding of biological materials \cite{ye2015supramolecular, patel2012review, sharma2014biomaterials}. 

Significant challenges still exist in accurately measuring mechanical properties of soft materials. Commonly, biological tissues that are excised exhibit altered properties upon testing \cite{nickerson2008rheological, zimberlin2010cavitation}. Further, in some instances biological material is geometrically contorted to meet standards of conventional testing methods, such as tensile testing, which complicates material characterization \cite{malone2018mechanical,krasokha2010mechanical}. Further, while indentation and rheometry methods for viscoelastic analysis have been utilized, \cite{balooch1998viscoelastic,zheng1999extraction, mahaffy2004quantitative, vanlandingham2005viscoelastic, hu2010using, budday2015mechanical} the understanding of material properties past the linear-elastic regime becomes limited  \cite{lin2009spherical, style2013surface}. This has led to a fragmented understanding of soft material properties and has limited insight into materials' nonlinear behaviors. 

To address these issues, a novel method known as Volume Controlled Cavity Expansion (VCCE) was developed \cite{raayai2019volume, raayai2019intimate}. VCCE offers an approach to measure the complex nonlinear responses of soft materials. It utilizes incompressible fluid, which is controllably injected, via a needle syringe system, locally into the material, while concurrently measuring pressure. This yields a detailed pressure-volume relation that captures the material's nonlinear response, enabling users to discern parameters such as age, hydration, or other tested conditions. Additionally, this measured result can be fit to relevant constitutive material models, such as the neo-Hookean \cite{gent1959internal}, Ogden \cite{ogden1972large}, or Fung \cite{chuong1983three} models, to determine mechanical properties of interest. This protocol has been successfully applied to a diverse array of materials including Polydimethylsiloxane (PDMS)\cite{chockalingam2021probing, raayai2019volume}, brain tissue \cite{mijailovic2021localized}, blood clots \cite{varner2023elasticity}, and liver \cite{nafo2021measuring}, showcasing its versatility and effectiveness. 

VCCE is the extension of the Needle Induced Cavitation Rheology (NICR) \cite{delbos2012cavity, fuentes2019using, chin2013cavitation, polio2018cross, zimberlin2010cavitation, cui2011cavitation, zimberlin2007cavitation, crosby2011blowing, zimberlin2010water, cui2011cavitation, delbos2012cavity, blumlein2017mechanical, kundu2009cavitation} method, which has represented a shift in how researchers have been able to evaluate soft materials. Though similar, the NICR method does not control volume and can only recover a single material parameter given its reliance on the cavitation instability. Through VCCE, the rate at which the fluid volume is introduced in the material is correlated to the measured resisting pressure and this pressure-volume relationship allows multiple material properties to be recovered. 

Despite its potential, VCCE has primarily been implemented using mechanical testing apparatuses such as universal testing machines \cite{varner2023elasticity, chockalingam2021probing, raayai2019intimate, raayai2019volume} which are used due to the high degree of precision required when displacing a plunger in a syringe-based system. Unfortunately, such systems pose barriers in terms of size (approximately 2m. x 1m. x 1m.), weight (700+ lbs.), power (220VAC) and cost (\$150,000+)\cite{instronTestingSystems}. These limitations have restricted VCCE's accessibility to researchers and industry professionals who have not only the financial resources to purchase these units, but also the facility and personnel resources to adequately dedicate space and time to become proficient for VCCE application. 

In response to this, we developed the first bench-top VCCE testing instrument, designed with Consumer Off-The-Shelf (COTS) components and accompanying open-access data-acquisition software, instructions, and data-processing code \cite{openVCCE}. This tool can offer immediate benefit to hospital environments where users face various equipment challenges \cite{zumba2023contribution, rawlinson1978space, tung2018factors, marin2015developing}. Similarly, this system offers simplicity in measurement and data collection which enable academic and industry researchers to discover soft material phenomena towards medical diagnostics, failure criteria, and mechanical properties.

\section{\label{sec:level2}{Volume-Controlled Cavity Expansion}}
Expansion of a fluid bubble in VCCE is performed using a syringe connected in-line to a pressure sensor. Upon inserting the needle into the material the plunger is controlled and the pressure-volume response is recovered. Through comparison with predictive models, this pressure-volume response is used to evaluate the material's nonlinear constitutive behavior. 

\begin{figure}[h!]
	\centering 
	\includegraphics[width=0.5\textwidth, angle=0]{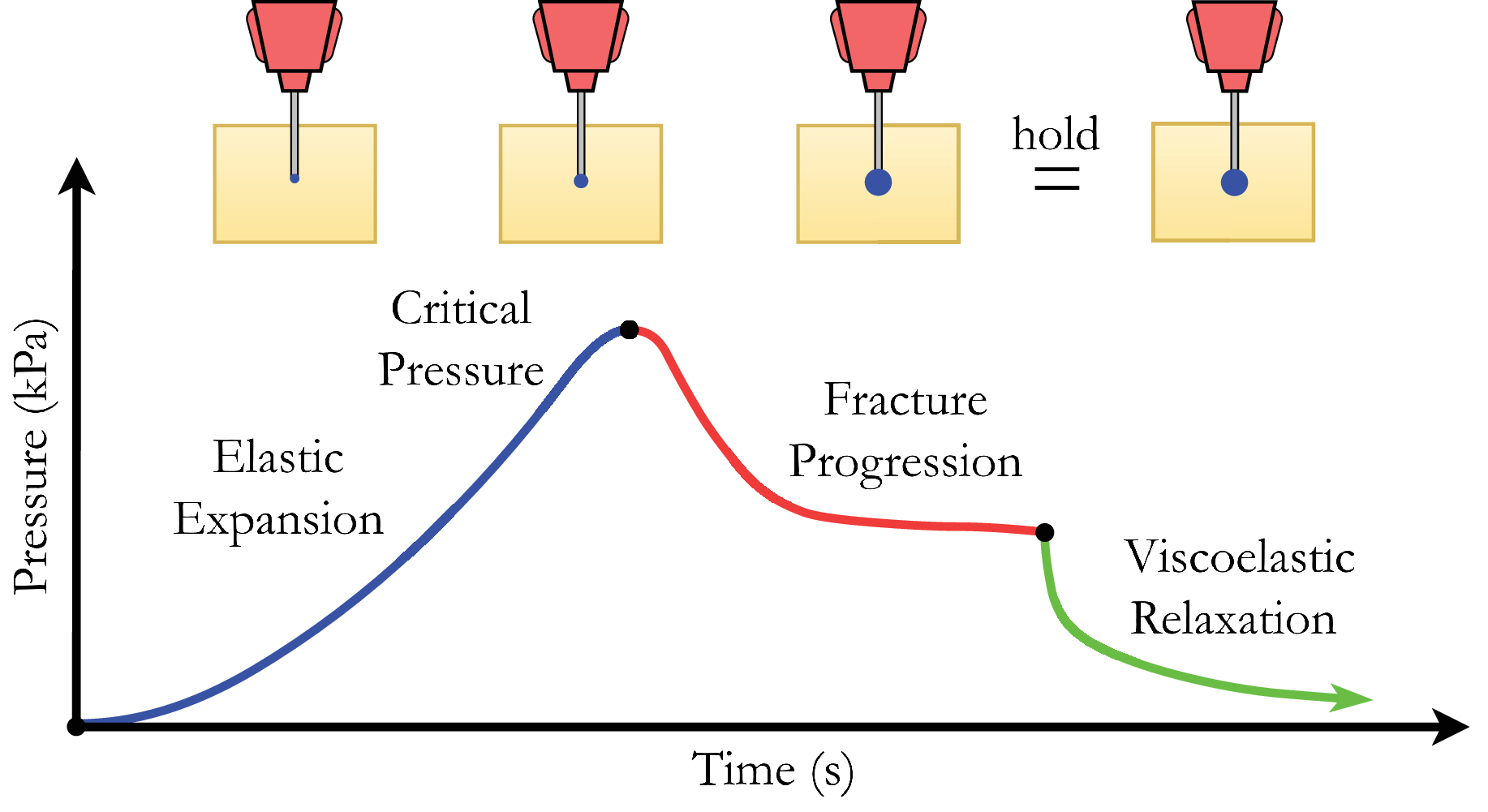}	
	\caption{Pressure versus time response for a constant fluid injection VCCE protocol via a needle-based system to recover the fully nonlinear response of a material under local loading conditions.} 
	\label{fig_1}%
\end{figure}

VCCE offers the flexibility of using various expansion and retraction scenarios. In this manuscript, to validate our instrument performance, we choose a specific protocol. We begin by expanding a bubble and continue expansion beyond the fracture limit at a constant volumetric flow rate. Then upon arriving at a prescribed volume, we cease expansion. This experimental approach enables us to capture various features in the nonlinear response of the material: the elastic expansion, the critical pressure at fracture, the fracture progression, and the viscoelastic relaxation, as illustrated in Fig. \ref{fig_1}.

\subsection{Verification via neo-Hookean Material Model}

\begin{figure}[h!]
	\centering 
	\includegraphics[width=0.20\textwidth, angle=0]{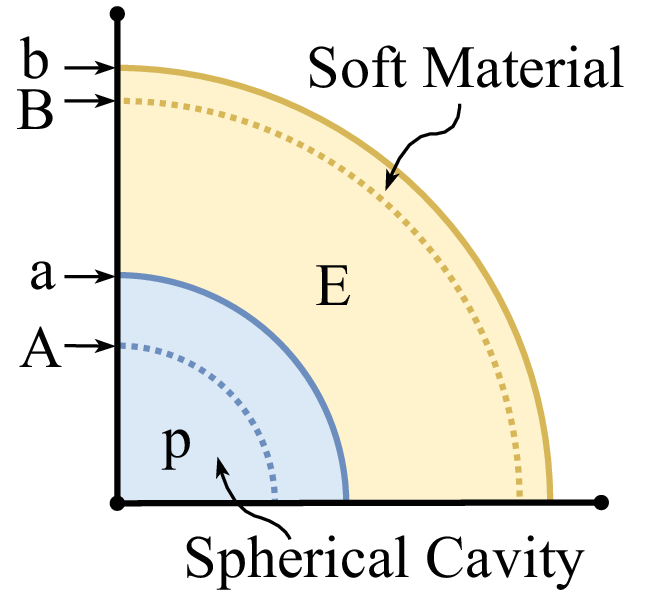}	
	\caption{An incompressible cavity expanding within an incompressible material of modulus, $E$, exerting pressure, $p$, at the cavity site. Original configurations, $A$ \& $B$, become deformed configurations, $a$ \& $b$.} 
	\label{fig_2_0}%
\end{figure}

For validation of our bench-top unit, we use PDMS, which has previously been shown \cite{raayai2019volume} to be well-characterized by the neo-Hookean model \cite{gent1959internal} in the quasi-static range. For this, we first define the circumferential stretch, $\lambda$: 
\begin{equation}
    \lambda = \frac{{a}}{A},
\end{equation}
where the effective radius of the spherical cavity, $a$, is divided by the effective initial defect radius, $A$. As shown in earlier studies\cite{raayai2019volume}, the initial defect, and subsequent expanding cavity, is well captured by the spherical assumption. We define the cavity, as an effective sphere, that expands to effective radius, $a$, in a soft material of undeformed radius, $B$, which under the influence of pressure, $p$, deforms to effective radius, $b$ (Fig. \ref{fig_2_0}).

Theoretical prediction of the cavity pressure as a function of stretch, assuming the cavity is small compared to the dimensions of the sample (i.e.  $B/A\to\infty$) and using the neo-Hookean model can be written in the form\cite{gent1959internal}
\begin{equation}
    \frac{p}{E}=\frac{5}{6}-\frac{2}{3\lambda}-\frac{1}{6\lambda^4}
    \label{eq_1}
\end{equation}
and  will be used in our instrumentation validation. Note that earlier studies\cite{raayai2019volume} have shown that the above formula applies also for finite bodies within reasonable  stretch  range (see Appendix Fig. \ref{fig_2}).

For verification, in this work we produce fluid cavities of effective final radius, $a$=1.3mm, with effective initial defect sizes, $A$$\approx$0.25mm, corresponding to $\lambda$$\approx$5. In order to neglect boundary effects in our analysis, and knowing empirically that PDMS fractures near $\lambda$$\approx$2, we need sample sizes of $B/A\geq$10, indicating $B\geq$2.5mm as sufficient for representing elastic data.
\section{\label{sec:level3} Instrumentation}
This instrument, constructed using readily available components and costing nominally \$5000 USD, streamlines the evaluation soft materials using the VCCE method, and makes VCCE testing accessible to users with limited technical backgrounds. The system can evaluate soft materials across elastic moduli ranging from 1 kPa to hundreds of kPa,  

This instrument decomposes VCCE into two primary subsystems: mechanical and electrical. Details on these subsystems and their components are provided herein.

\subsection{Bench-top Overview}

\begin{figure*}
	\centering 
	\includegraphics[width=\textwidth, angle=0]{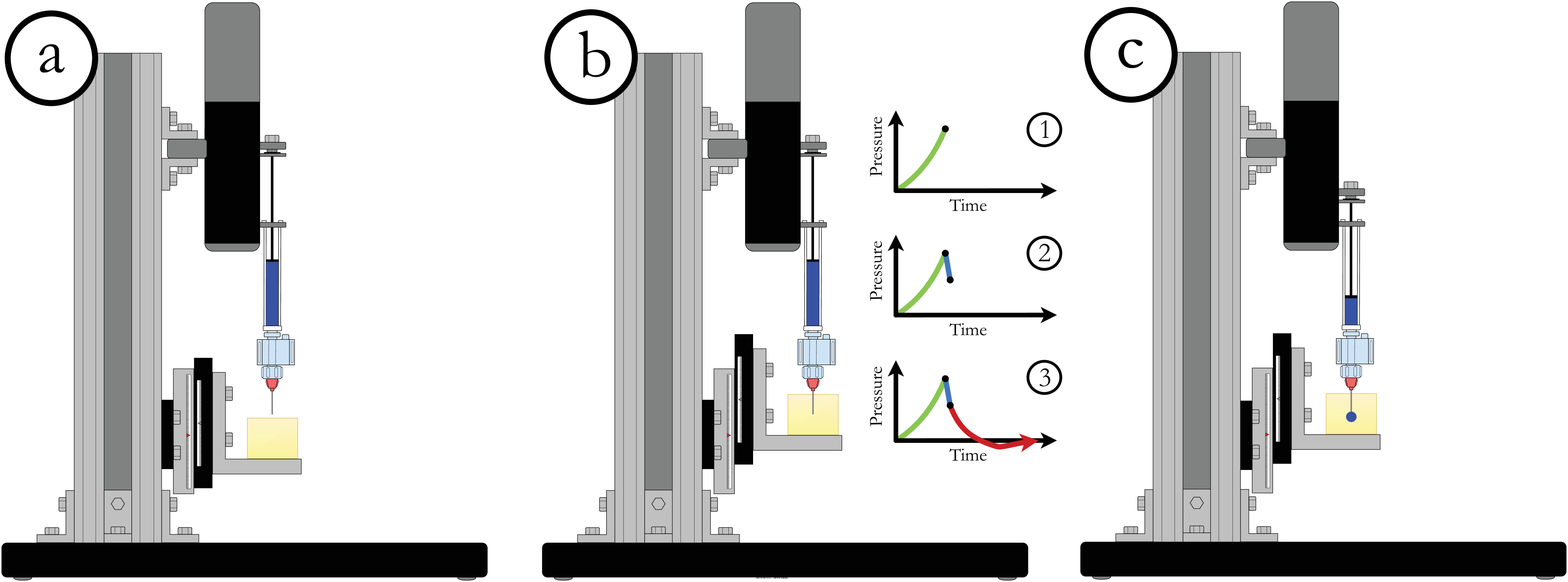}	
	\caption{VCCE testing is performed in three stages: a) loading a sample on the translation stage, b) raising the translation stage into the syringe assembly and zeroing pressure, and lastly, c) controlling the rate of cavity creation within the material. Specifically, the green line (b.1) represents the needle being inserted into the material, depressing the surface of a material, but the needle has not been penetrated the material surface yet. The blue line (b.2) represents when the needle has penetrated the material surface and a near instantaneous drop in pressure is seen. The needle is then further plunged a distance into the material. Afterwards the needle is slowly retracted, represented by the red line (b.3) until the pressure reaches approximately zero.}
	\label{fig_0}%
\end{figure*}

\begin{figure}[h!]
	\centering 
	\includegraphics[width=0.5\textwidth, angle=0]{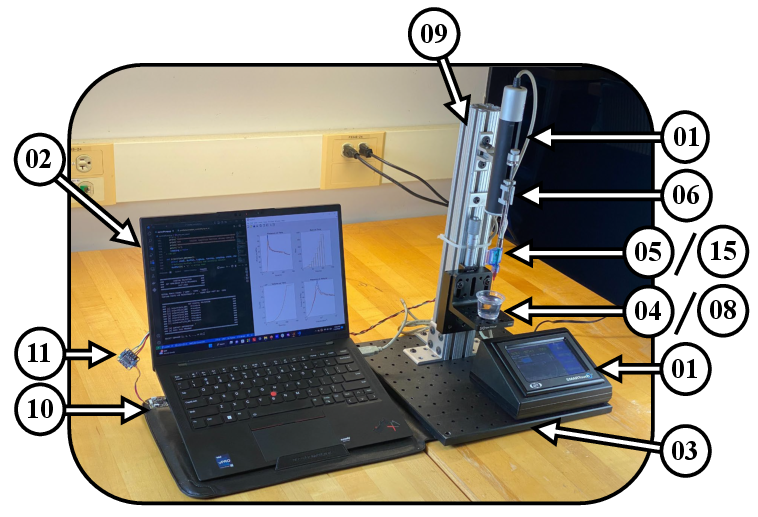}	
	\caption{Assembled bench-top VCCE unit with major components listed in Table \ref{table_1}.} 
	\label{fig_3}%
\end{figure}

\begin{table}
\caption{Component List}
\begin{ruledtabular}
\begin{tabular}{l c c} 
{{[\#\#]}} & {Component Name} & {QTY} \\ \hline 
{[01]} & WPI UMP3T \& Controller w/ USB Cable & 1 \\
{[02]} & Windows Laptop Computer & 1 \\ 
{[03]} & Optical Breadboard Baseplate & 1 \\ 
{[04]} & Translation Stage with Standard Micrometer & 1 \\ 
{[05]} & PendoTech Pressure Sensor & 1 \\ 
{[06]} & Hamilton 10$\mu$L Syringe & 1 \\ 
{[07]} & Aluminum Bolt-Together Corner-Bracket & 5 \\ 
{[08]} & Right-Angle Bracket with Counterbored Slots & 1 \\  
{[09]} & T-slotted Framing Rail & 1 \\ 
{[10]} & RP2040 USB Key & 1 \\ 
{[11]} & Nuvoton NAU7802 24-Bit ADC & 1 \\ 
{[12]} & 1/4-20 Socket Head Screws & 18 \\ 
{[13]} & T-Slotted Framing Fasteners & 14 \\ 
{[14]} & QT-to-QT Cable & 1 \\ 
{[15]} & 25G Luer-lock needle connection & 1 \\ 
{[16]} & Working Fluid & 400$\mu$L \\ 
\end{tabular}
\end{ruledtabular}
\label{table_1}
\end{table}

The bench-top system, as depicted in Figs. \ref{fig_0} and \ref{fig_3} and with components shown in Table \ref{table_1}, is designed to work with laptop computers equipped with a minimum of two USB Type-A ports. Specifically, these ports are allocated for distinct functions: one for data acquisition - particularly pressure data - and the other for managing the controls hardware. Data acquisition is facilitated through a PendoTech pressure sensor, connected to a 10$\mu$L syringe. This sensor is sequentially connected to an analog-to-digital converter and then a microcontroller-based USB key. The USB key, once inserted into the laptop computer, is accessed and controlled via Python scripting in Visual Studio Code, enabling data collection. 

The selected hardware components include commercially available syringe microinjector systems from World Precision Instruments (WPI), specifically the UMP3T and MICROTOUCH 2T models. These systems are directly connected to the laptop computer through USB and interrogated via the same Python scripting that handles data collection. Simultaneous data acquisition through two USB ports enables the synchronization of pressure readings with fluid injected into the soft material. 

Before initiating data collection, the bench-top frame requires assembly. The microinjector subassembly is mounted onto a T-slotted framing structure, stabilized by an optical breadboard baseplate. Positioned directly beneath the microinjector subassembly and affixed to the T-slotted frame is a rack-and-pinion translation stage, outfitted with an embedded micrometer acting as the sample stage. This stage raises the soft material and inserts the needle into the sample in preparation for a VCCE test. The process of elevating the stage to introduce the sample into the needle is designed to create an initial cavity within the material, establishing a consistent and repeatable zero-condition for subsequent material analysis. Once inserted, a VCCE test may be performed and is outlined in greater detail in Section \ref{sec:level4}. 


\subsection{\label{sec:level3_1}Evaluating Instrumentation Compliance}
Any compliance in the mechanical subsystems during operation can lead to perceived losses in volume. Such losses, can reduce accuracy of the inferred volume and subsequently the pressure-volume response used to assess nonlinear material behaviors. Additionally, the stored elastic energy in the mechanical subsystems can lead to an unstable pressure drop if the material exceeds the critical pressure. Hence, it is essential to evaluate compliance of components within the mechanical subsystem and to ensure material behavior pressure drops are not due to system compliance, but instead reflective of mechanical behavior.

First, we begin with examination of the instrument's compliance. When considering compliance within our instrument, we restrict our attention to the operating state when the liquid cavity is introduced into the soft material. Evaluating forces that arise in operation, components directly in contact with the pressurized working fluid, will undergo predictable deformations that should be evaluated under maximum operational conditions. We have the capability to characterize our system up to the sensor's maximum pressure of $\sim$500kPa; however, peak pressures of soft materials rarely reach such levels before fracture, even if their elastic modulus surpasses that figure. Therefore, we focus our analysis on local peak pressures up to 100kPa.

We focus our attention to components that are likely to exhibit potentially significant compliance under the anticipated local peak pressures up to 100kPa, including the working fluid, the pressure sensor's Polycarbonate housing, its compliant dielectric silicone Micro-Electro-Mechanical systems (MEMS) diaphragm, and the Polypropylene 25G Luer-lock needle connection. Each component is evaluated under worst-case loading conditions to highlight the role of compliance during measurement as material stiffness increases. 

\begin{figure}[h!]
	\centering 
	\includegraphics[width=0.5\textwidth, angle=0]{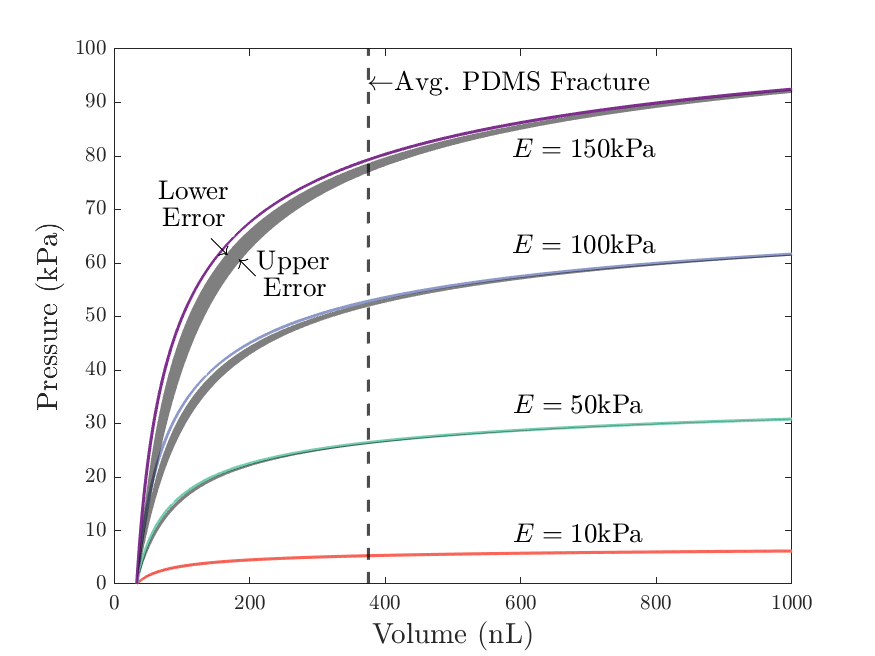}	
	\caption{Pressure versus volume response of a neo-Hookean material with varying stiffness. Gray regions denote the role of component compliance where \textit{Lower Error} (Eq. \ref{eq_20_1}) reflects water compressibility and \textit{Upper Error} (Eq.\ref{uppererror}) reflects select component compliances along with water compressibility. As material stiffness increases, accuracy becomes compromised, but measurement remains repeatable.} 
	\label{fig_9comp}%
\end{figure}

A component which contributes towards measurement compliance is the working fluid -- water. Different fluids may be used during VCCE testing and in this analysis we highlight water as a worse-case scenario since it is more compressible than another primary alternative -- oil. Utilizing the bulk modulus, $K$, of salinated water at 2.22GPa, and an initial fluid volume, $V$, of 400$\mu$L under pressure, $P$, the compressibility of the working fluid, $\Delta{V_w}$, is calculated as follows:
\begin{equation}
    Lower Error: \quad \Delta{V_w} = -\frac{P{V}}{K}.
    \label{eq_20_1}
\end{equation}
\indent When translated to the neo-Hookean result (Eq. \ref{eq_1}), as shown in Fig. \ref{fig_9comp}, the \textit{Lower Error} (Eq. \ref{eq_20_1}) - representing the compliance of water under various pressure conditions - becomes increasingly noticeable as the pressure and, consequently, the sample stiffness, increases. Neglecting fluid compression at higher recorded pressures can lead to an underestimation of the volume injected into the soft material, thereby affecting the accuracy of reported material properties.


In our system the pressure sensing is based on the deformation of a compliant dielectric silicone membrane. We estimate the deflection of this thin disc using Plate Theory \cite{timoshenko1959theory} where the deflection, $w(\zeta)$, of a thin disc of height, $h\approx0.1$mm, radius, $r\approx1.0$mm, with uniform distribution, $P_s$, can be tracked from the center of the disc as a function of radial distance, $\zeta$, follows: 
\begin{equation}
    w(\zeta) = -\frac{{P_s}}{64D}\left(r^2-\zeta^2\right)^2,
    \label{eq_14}
\end{equation}
\noindent where the dielectric silicone membrane has properties of $E\approx5.0$MPa, $\nu\approx0.48$, and, $D$, is the corresponding flexural rigidity: 
\begin{equation}
    D=\frac{Eh^3}{12(1-\nu^2)}.
    \label{eq_12}
\end{equation}
\indent Integrating yields the following equation describing the volumetric loss due to the MEMS pressure sensor: 
\begin{equation}
    \Delta{V_{m}}=-\frac{\pi{\zeta^6{P_s}}}{192D}.
    \label{eq_15}
\end{equation}
\indent Next we evaluate the Polycarbonate pressure sensor housing and Polypropylene Luer-lock needle connection. For this, we analogize these components as thick-walled pressure vessels with Polycarbonate having properties $E\approx2.4$GPa and $\nu\approx0.35$, and Polypropylene having estimated properties $E\approx1.5$GPa and $\nu\approx0.42$. Both components exhibit varying degrees of longitudinal ribbing and structural reinforcements, which impacts their effective stiffness and radii. To account for this, we assume that the reinforcements increase the effective outer radius by 50\% in our calculations. Therefore the Polycarbonate pressure sensor housing's geometry is, outer radius, $r_o\approx3.24$mm, inner radius, $r_i\approx1.45$mm, with vessel height, $h_s\approx25.4$mm, and the Polypropolene Luer-lock needle connection's geometry is $r_o\approx4.65$mm, $r_i\approx2.30$mm, with $h_s\approx3.18$mm. The vessel is then subjected to constant external pressure, $P_o$, and increasing internal pressure, $P_i$, which after defining the ratio of the outer radius over the inner radius as $\alpha=r_o/r_i$ results in the change in the vessel's inner radius\cite{anand2023introduction}:
\begin{equation}
        \frac{\Delta{r_{r_i}}}{r_i} = \frac{{1+\nu}}{E}\left[\frac{P_i(1-2\nu+\alpha^2)-2P_o\alpha^2(1-\nu)}{\alpha^2-1}\right]-\nu\epsilon_o,
\end{equation}
\noindent where, $\epsilon_o$, the axial strain, represents a capped cylinder characterized by: 
\begin{equation}
    \epsilon_o = \frac{1-2\nu}{E} \left( \frac{P_i{r_i}^2-P_o{r_o}^2}{r_o^2-r_i^2} \right).
\end{equation}

This is subsequently expressed as a change in vessel volume, $\Delta{V_v}$, relevant to the VCCE measurement:
\begin{equation}
    \Delta{V_v}=\pi h_{s} \left( \Delta{r_{r_i}}^2 + 2 \Delta{r_{r_i}}r_i \right).
    \label{eq_20}
\end{equation}


In Fig. \ref{fig_9comp}, the \textit{Upper Error} (Eq. \ref{uppererror}) represents the total summed volumetric losses due to water compliance, $\Delta{V}_w$, MEMS compliance, $\Delta{V}_m$, compliance from the pressure sensor housing, $\Delta{V}_{v_{1}}$, and Luer-lock needle connection, $\Delta{V}_{v_{2}}$.
\begin{equation}
    UpperError: \quad \Delta{V}_w + \Delta{V}_m + \Delta{V}_{v_{1}} + \Delta{V}_{v_{2}}
    \label{uppererror}
\end{equation}
\indent Fig. \ref{fig_9comp} shows the neo-Hookean result (Eq. \ref{eq_1}) without any errors compared to the neo-Hookean result with anticipated compliance errors. As stated previously, the role of system compliance becomes noticeable for higher pressures. Nevertheless, it is important to note that this system compliance is predictable and repeatable; hence, while it may influence the measurement accuracy, it will not impact the measurement precision.

\subsection{\label{sec:level3_99}Motor Stepping and Precision}
\begin{figure}[h!]
    \centering 
	\includegraphics[width=0.5\textwidth, angle=0]{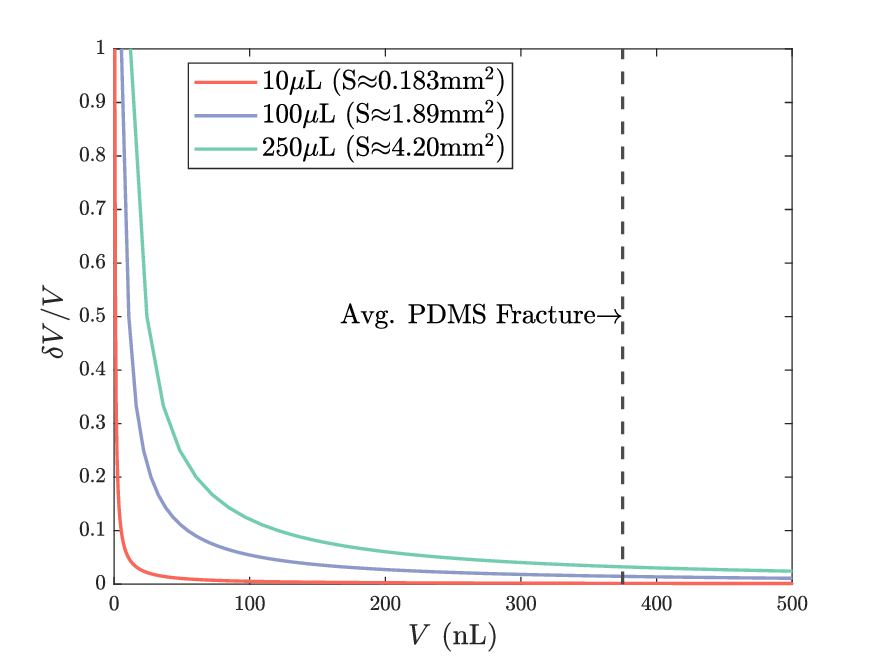}	
	\caption{Precision of volume control throughout the volume expansion process is shown for three syringes of different cross-sectional areas.} 
	\label{fig_step_res}%
\end{figure}
In VCCE, selecting an appropriate stepper motor resolution and syringe diameter are both important for precise capture of  the elastic branch of the expansion.  To estimate the error in fluid volume control, we write the fluid transfer resolution as:
\begin{equation}
    \delta{V}=S\Delta{X}_n,
    \label{step_res1}
\end{equation}
where $S$ is the syringe cross-sectional area, and $\Delta{X}_n$ is the motor stepping resolution, which for our system is approximately 2.8$\mu$m. The transferred volume, ${V}$, can thus be written as a function of the number motor of steps, $N$:
\begin{equation}
    V=N\delta{V},    
    \label{step_res2}
\end{equation} such that the instantaneous error in volume control is $\delta V/V$.

\indent As shown in Fig. \ref{fig_step_res} the precision in volume control increases as expansion proceeds (i.e. the error decreases). Among the three standard glass syringes that were considered, the 10$\mu$L syringe ($S \approx 0.183$mm$^2$) provides high precision with negligible error throughout the range of interest. Therefore, the 10$\mu$L syringe is selected for further material analysis and instrument validation.


\subsection{\label{sec:level3_4}The Electrical Subsystem}
\subsubsection{\label{sec:level3_4_1}Pressure Sensing Configuration}
The initial component of the electrical subsystem incorporates a PendoTech single-use pressure sensor, designed to measure both static and dynamic pressures of gases and liquids. The sensor operates via a Wheatstone bridge configuration \cite{hoffmann1974applying}, where pressure is recorded through balancing a resistor bridge, and correlating the change in resistance at equilibrium, to a measure of pressure.

The sensor is capable of measuring pressures ranging -11.5 to 75 psi. The sensor's accuracy is delineated as follows: $\pm$2\% for the 0 to 6 psi range, $\pm$3\% for the 6 to 30 psi range, and $\pm$5\% for the 30 to 60 psi range.

\subsubsection{\label{sec:level3_4_2}Signal Processing and Power Supply}
Signal conversion from the MEMS-based PendoTech pressure sensor to a computer-interpretable format is achieved using Nuvoton's NAU7802, a 24-Bit Analog-to-Digital Converter (ADC) specifically designed for Wheatstone bridge sensors. The digital output is interfaced with a RP2040 microcontroller embedded USB Key, which facilitates data transmission to the host computer via serial communication. Power to the pressure sensor is provided directly from the host computer's USB port, with voltage regulation provided by the RP2040 microcontroller, ensuring compatibility with the Inter-Integrated Circuit (I2C) communication protocol. An unregulated 5.00VDC from a USB port is first regulated down to 3.30VDC within the RP2040, and then further reduced to 3.00VDC for the pressure sensor. The received signal is converted to a measure of pressure via the following conversion 0.2584mV/V/psi where the supply voltage used for calculation is 3.00VDC.

\indent Further calibration could be necessary if the temperature of the system were varying in time. To this end, we begin all tests by writing to the Nuvoton ADC 0x11 hexadecimal register and enabling temperature sensing. We record ambient temperature, then automatically archive this value into our master test logs for each test. If significant temperature variation between tests were observed a correction factor of 0.04kPa/C could be applied; however, that action was not taken in this study. 

\section{\label{sec:level4}Validation Testing}
\subsection{\label{sec:level4_1}PDMS Fabrication Procedure}
PDMS (DOW SYLGARD 184) samples were prepared with base-to-curing agent ratios of 43:1, 45:1, and 47:1, where a higher ratio corresponds to  lower sample stiffness\cite{raayai2019volume}. Each ratio mixture was separately processed in 150mL resin containers, designed for compatibility with the Thinky SR-500 planetary mixer, creating a homogeneous blend of components.

\indent The PDMS mixtures were subjected to a two-phase mixing regimen within the planetary mixer. Following the mixing process, the homogenized PDMS was immediately transferred to a vacuum chamber. The degassing stage removes entrapped air within the PDMS. Subsequently, the degassed mixture was poured into disposable plastic 2oz cups.

The filled 2oz containers were then placed in a curing oven set to 100C for a duration of two hours. Samples were then allowed to rest at room temperature for eight days post-curing. This additional resting period ensured any residual cross-linking reactions were complete, stabilizing the material's properties while minimizing potential long-term stiffening effects.

\subsection{\label{sec:level4_3}Testing Protocol}

Data acquisition and motor control in our system are managed through a Python script. This script integrates both controls and data acquisition.

For new users, it's necessary to download and install a specific firmware version for the RP2040 USB key. This firmware and other supporting materials can be found in the Nonlinear Solid Mechanics Group GitHub\cite{openVCCE}.

\begin{figure}[h!]
	\centering 
	\includegraphics[width=0.5\textwidth, angle=0]{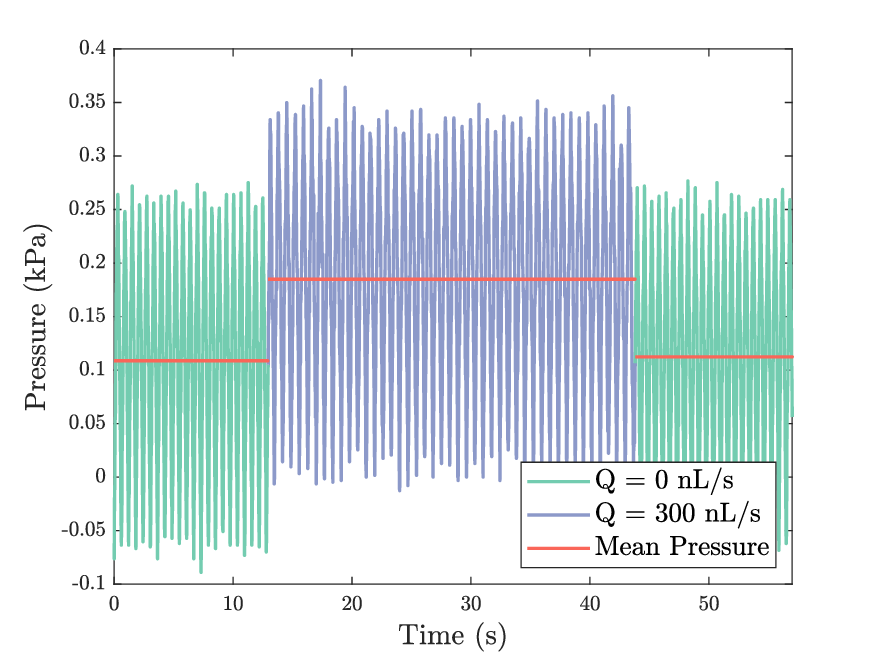}	
	\caption{Pressure versus time response for a water calibration of a 10$\mu$L syringe at $Q$=300nL/s.} 
	\label{fig_4_1}%
\end{figure}

Before experimenting on soft materials, it is essential to perform a dynamic calibration test using working fluid. This test accounts for potential losses due to fluid movement captured in the pressure response (Fig. \ref{fig_4_1}). These losses are averaged and subtracted from the final soft material datasets. This involves immersing the syringe needle in the working fluid. We then selected a constant volumetric flow rate, $Q$, of 300nL/s for a final cavity radius, $a$, of 1.3mm, calculated from the total volume of fluid, $V_T$, used: 

\begin{equation}
    V_T=\frac{4}{3}\pi{a^3}
    \label{eq_Vt}
\end{equation}

Once the water calibration is completed, the testing of soft materials can commence. Microinjector systems, characterized by their bi-directional motor functionality, necessitate a specific minimum engagement distance between the syringe plunger carrier and the threaded rod that connects to the stepper motor. For the WPI UMP3T setup, this distance is approximately 100$\mu$m. Utilizing a 10$\mu$L syring, the system requires an advancement of approximately 34.5 steps to begin testing, with each step dispensing approximately 0.6nL, totaling an approximate 20nL of fluid expulsion before the initiation of any tests. On code startup, the motor is programmed to move approximately 50nL at a delivery rate of $Q$=100nL/s to overcome this engagement distance. Following this preliminary step, the ADC is calibrated to zero.

The needle insertion process involves real-time pressure feedback displayed in the Visual Studio Code terminal. The schematic representation of the pressure response as a needle inserts into the material, prior to testing, is seen in the subsets of Fig. \ref{fig_0}. 
Completing the insertion procedure, an initial defect is created, where the size of the initial defect can be influenced by the amount of retraction. Specifically in this work, to minimize residual stresses that emerge in the penetration process we follow the protocol identified in Fig. \ref{fig_0}, which defines the final insertion depth to zero the measured pressure. For the tests conducted in our validation testing, we provide the corresponding insertion depths in Table \ref{table_5} (see Appendix).  Once the insertion procedure is completed the testing setup is prepared for fluid injection.

After completing the needle insertion procedure, the user enters the main VCCE testing protocol. Tests were performed with $Q$=300nL/s until reaching a desired cavity radius of 1.3mm, equivalent to a total injection volume, $V_T$, of 9202.8nL. The entire testing sequence was executed 21 times, in direct succession, to ensure precision and accuracy of the instrument. 

\section{\label{sec:level5}Results \& Analysis}
\begin{figure}[h!]
	\centering 
	\includegraphics[width=0.5\textwidth, angle=0]{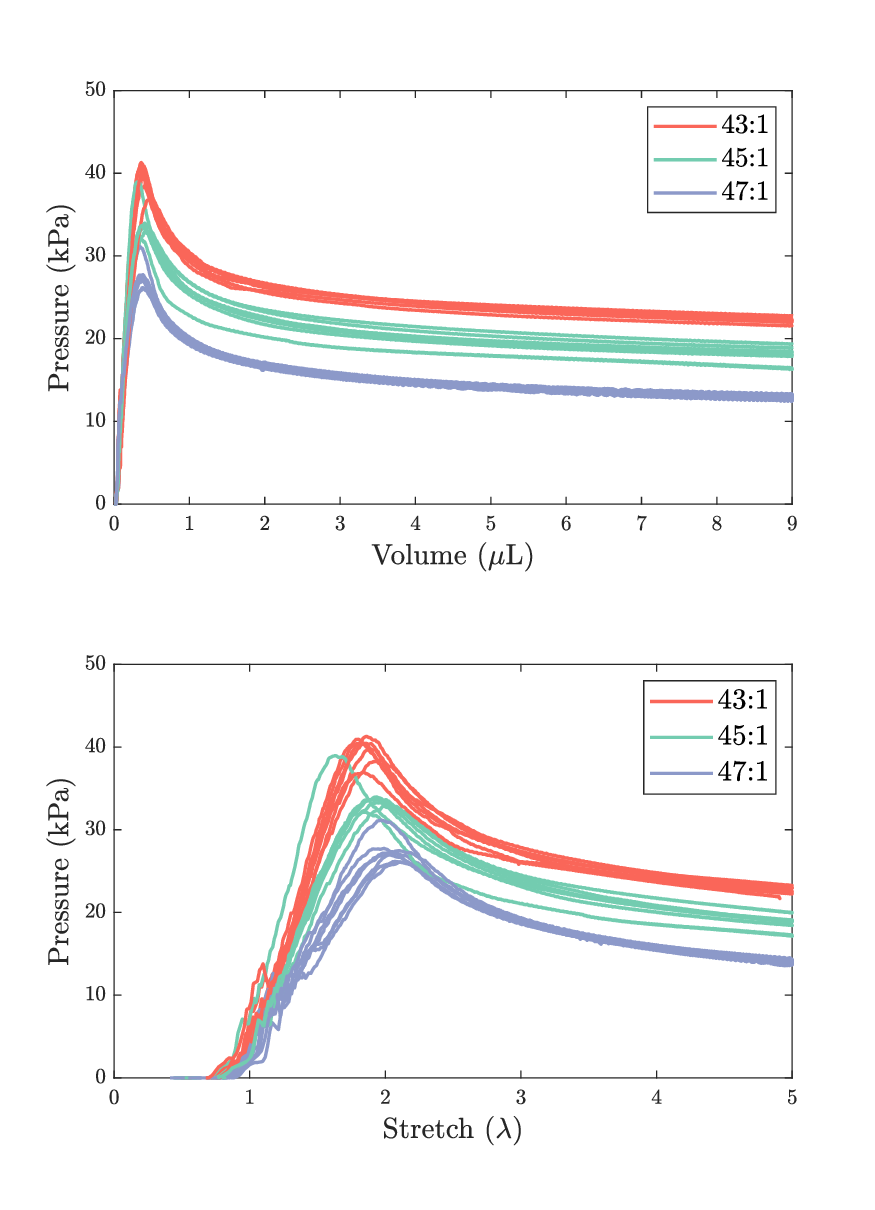}	
	\caption{Calibrated material response curves for 21 successive tests of testing PDMS of ratios 43:1 (red), 45:1 (green), and 47:1 (blue) for both $p-V$ (top) response curves and $p-\lambda$ (bottom) curves.} 
	\label{fig_6}%
\end{figure}

Data was collected from PDMS base:curing agent ratios of 43:1, 45:1, and 47:1 (Fig. \ref{fig_6}). Each ratio group was represented by 7 distinct tests, designed to elucidate key mechanical properties from the elastic expansion, critical pressure, and fracture progression (Fig. \ref{fig_1}). Comprehensive datasets, encompassing all collected data and further observations on viscoelastic relaxation, are accessible via the GitHub\cite{openVCCE}.

The samples collected exhibit clear trends prior to constitutive fitting (Fig. \ref{fig_6}). The maximum of the dataset, $P_c$, are observed to decrease with an increase in the base:curing agent ratio in PDMS, suggesting a correlation between material stiffness and critical pressure preceding fracture onset. Specifically, the average critical pressures for ratios of 43:1, 45:1, and 47:1 are 39.76$\pm1.54$kPa, 34.12$\pm2.23$kPa, and 27.57$\pm1.70$kPa, respectively. Moreover, the PDMS material is noted to fracture at volumes, $V_c$, measured in nanoliter, with the 43:1 ratio fracturing at an average of 390.6$\pm45.3$nL, 45:1 at 374.5$\pm38.9$nL, and 47:1 at 366.1$\pm19.1$nL. 



By employing numerical integration of the resulting pressure, $p$, up to $V_c$, we can calculate the work exerted by the expanding cavity on the soft material until fracture:
\begin{equation}
    U_c = \int_{0}^{V_c} p \, dV.
    \label{eq_30}
\end{equation}
\indent Additionally, we integrate the entire nonlinear response, up until the total volume injected, $V_T$, to find the work exerted by the expanding cavity on the soft material during the entire test: 
\begin{equation}
    U_s = \int_{0}^{V_T} p \, dV.
    \label{eq_31}
\end{equation}

\begin{figure}[h!]
	\centering 
	\includegraphics[width=0.5\textwidth, angle=0]{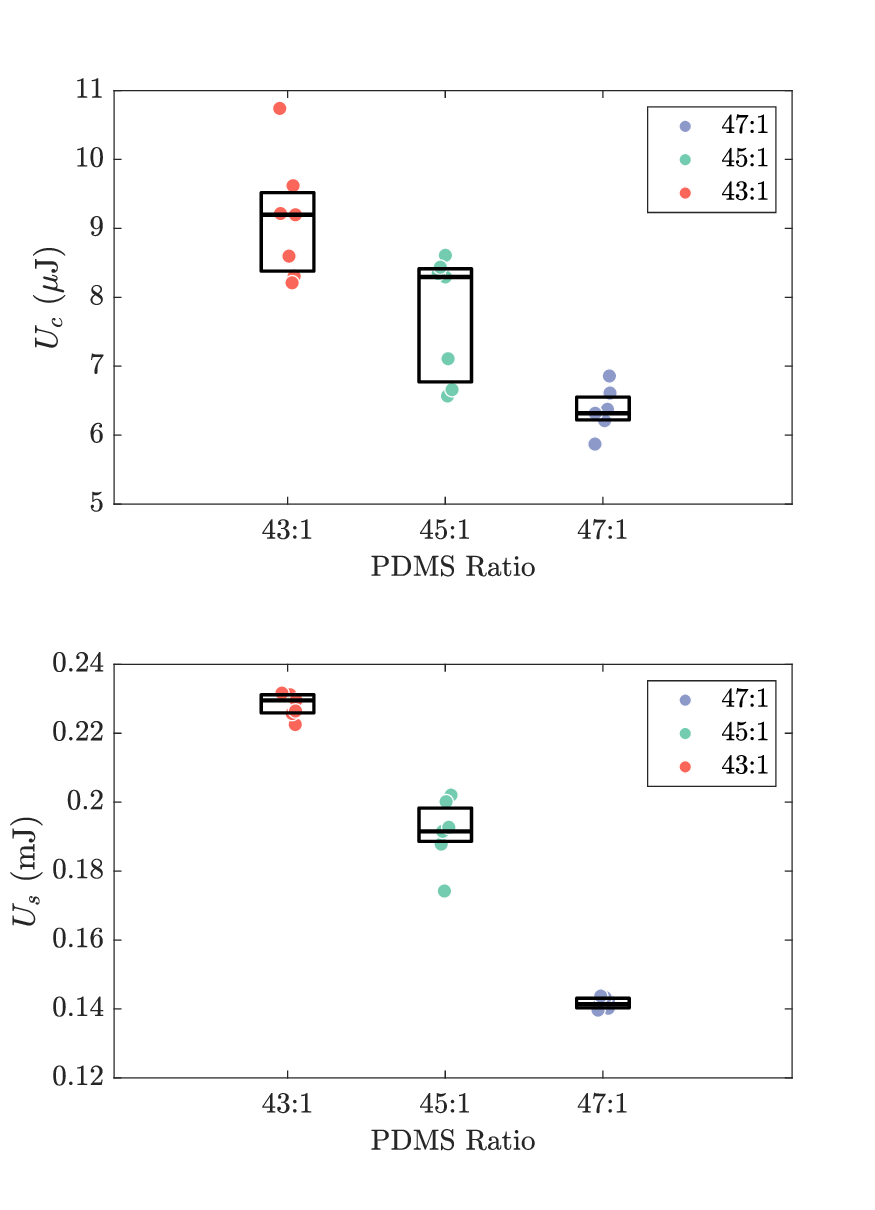}	
	\caption{Work from the fluid cavity expanding on the soft material up until fracture, $U_c$ (top), for the three different PDMS base:curing agent ratios. Additionally, the work exerted by the expanding fluid cavity on the soft material during the entire test, $U_s$ (bottom), for the three different PDMS base:curing agent ratios can be seen.} 
	\label{fig_8}%
\end{figure}

Further, showing the measured distributions of $U_c$ and $U_s$ for our 43:1, 45:1 and 47:1 samples (Fig. \ref{fig_8}), we see the instrument performs well in being able to differentiate materials based on their ability to sustain energy. Focusing on the region before fracture, the system effectively groups energy, and trends can be observed as 43:1 yielded, $U_c$, equal to 9.13$\pm$0.88$\mu$J, 45:1 equal to 7.72$\pm$0.90$\mu$J and 47:1 equal to 6.36$\pm$0.31$\mu$J. Focusing on the total volume of fluid injected, 43:1 yielded, $U_s$, equal to 0.23$\pm$0.003mJ, 45:1 equal to 0.19$\pm$0.009mJ, and 47:1 equal to 0.14$\pm$0.002mJ. 

\begin{figure}
	\centering 
	\includegraphics[width=0.5\textwidth, angle=0]{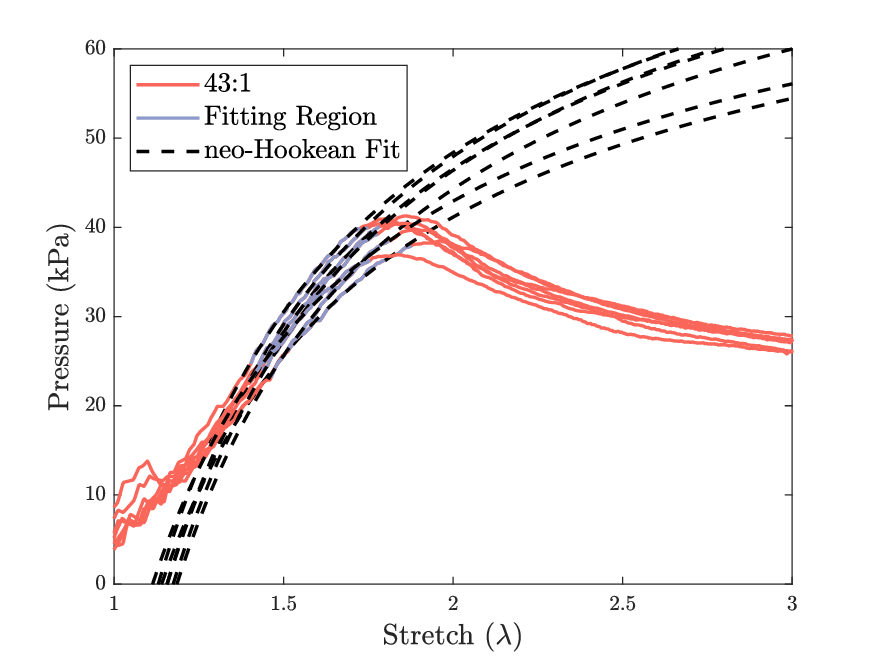}	
	\caption{Highlighting the neo-Hookean fitting procedure for 43:1 PDMS samples. The fitting region, illustrates where the spherical cavity assumption holds true and is a region of high elastic energy.} 
	\label{fig_10}%
\end{figure}

The data was further processed using the neo-Hookean constitutive model (Eq. \ref{eq_1}) to attain a measure of the elastic modulus for each PDMS ratio tested. Note that the modulus reported throughout is an instantaneous elastic modulus, $E$, due to initially high stretch rate expansions\cite{chockalingam2021probing}. Initially, the dynamic pressure response obtained from the infusion of the working fluid solution into another cup of working fluid at $Q$=300nL/s was subtracted from the PDMS data. Subsequently, the nonlinear results for PDMS were truncated between the geometric values of the effective radius of the 25G needle (257.5$\mu$m) and the radius of critical pressure for each sample. Within this range, the steepest slope was identified and extrapolated to identify the abscissa, which corresponds to an estimated initial defect size, $A$. We then fit a consistent region of the raw pressure-volume data, based on the previously identified steepest slope, to the neo-Hookean constitutive model (Eq. \ref{eq_1}), which can be seen in Fig. \ref{fig_10}. 

As seen in Fig. \ref{fig_10}, the neo-Hookean model (Eq. \ref{eq_1}), which describes the elastic properties of materials, falls short in capturing hyperelastic material fracture when subjected to large deformations. Thus, the utility of this model is primarily confined to providing insights into the elastic characteristics of materials.

To accurately determine the elastic modulus, $E$, alongside a fitted initial defect size, $A$, and the corresponding stretch value, $\lambda$, the least-squares method was used.

\begin{table}
\caption{Assembled experimental information of neo-Hookean fits, peak recorded pressures, and values of energy. Note that $V_o$ is calculated from the ascertained initial defect, $A$. The initial defect is translated to the initial volume, $V_o$, via Eq. \ref{eq_Vt} through substituting $A$ for $a$.}
\centering 
\begin{ruledtabular}
\begin{tabular}{l c c c c c c c} 
 Sample & $V_o$ (nL) & $V_c$ (nL)& $\lambda_c$ & $P_c$ (kPa)& $E$ (kPa)& {$U_c$} ($\mu$J) & {$U_s$} (mJ) \\ \hline
 43-1 & 60.8 & 354.0 & 1.80 & 40.9 & 110.1 & 8.31 & 0.231 \\
 43-2 & 53.0 & 399.7 & 1.96 & 38.5 & 94.5 & 9.62 & 0.223 \\
 43-3 & 58.2 & 353.9 & 1.83 & 40.4 & 111.7 & 8.21 & 0.226 \\
 43-4 & 55.8 & 359.9 & 1.86 & 41.3 & 108.5 & 8.60 & 0.231 \\
 43-5 & 58.2 & 393.5 & 1.89 & 39.8 & 106.0 & 9.20 & 0.232 \\
 43-6 & 60.9 & 389.7 & 1.86 & 40.5 & 109.9 & 9.22 & 0.230 \\
 43-7 & 78.3 & 483.3 & 1.83 & 36.9 & 96.3 & 10.74 & 0.226 \\ \hline
 45-1 & 56.5 & 411.2 & 1.94 & 34.0 & 83.5 & 8.61 & 0.202 \\
 45-2 & 51.0 & 408.2 & 2.00 & 32.9 & 79.6 & 8.30 & 0.191 \\
 45-3 & 54.7 & 356.5 & 1.87 & 33.6 & 85.0 & 7.11 & 0.188 \\
 45-4 & 50.5 & 392.2 & 1.98 & 33.4 & 79.7 & 8.35 & 0.192 \\
 45-5 & 71.0 & 306.2 & 1.63 & 39.0 & 114.0 & 6.57 & 0.200 \\
 45-6 & 56.1 & 348.1 & 1.84 & 32.1 & 82.8 & 6.66 & 0.174 \\
 45-7 & 54.1 & 398.8 & 1.95 & 33.8 & 83.9 & 8.44 & 0.193 \\ \hline
 47-1 & 45.5 & 372.0 & 2.01 & 27.0 & 65.1 & 6.25 & 0.141 \\
 47-2 & 42.8 & 379.8 & 2.07 & 26.2 & 64.3 & 6.37 & 0.140 \\
 47-3 & 40.6 & 376.8 & 2.10 & 26.1 & 64.9 & 6.21 & 0.141 \\
 47-4 & 43.9 & 376.2& 2.05 & 27.4 & 66.4 & 6.61 & 0.142 \\
 47-5 & 48.3 & 380.8 & 1.99 & 27.8 & 68.2 & 6.86 & 0.143 \\
 47-6 & 44.6 & 339.3 & 1.97 & 31.2 & 84.5 & 6.32 & 0.144 \\
 47-7 & 36.4 & 337.7 & 2.10 & 27.5 & 68.5 & 5.87 & 0.140 \\
\end{tabular}
\end{ruledtabular}
\label{table_4}
\end{table}

Fitted average values for the initial defect sizes, $A$, exhibit a downward trend with the values for 43:1 ratio being 0.243$\pm$0.010mm, for 45:1 being 0.237$\pm$0.009mm, and for 47:1 being 0.217$\pm$0.006mm. This observed trend suggests that softer materials may produce smaller initial defects, reflecting ease of defect formation. Similarly, critical stretch values, $\lambda_c$, indicative of the stretch at which fracture is presumed to occur, exhibit an upward trend as the material softens. Specifically, the average $\lambda_c$ values are 1.86$\pm$0.05 for the 43:1 ratio, 1.89$\pm$0.13 for the 45:1 ratio, and 2.04$\pm$0.05 for the 47:1 ratio. This pattern aligns with the expectation that softer materials can undergo greater deformation before fracture.

\begin{figure}
	\centering 
	\includegraphics[width=0.5\textwidth, angle=0]{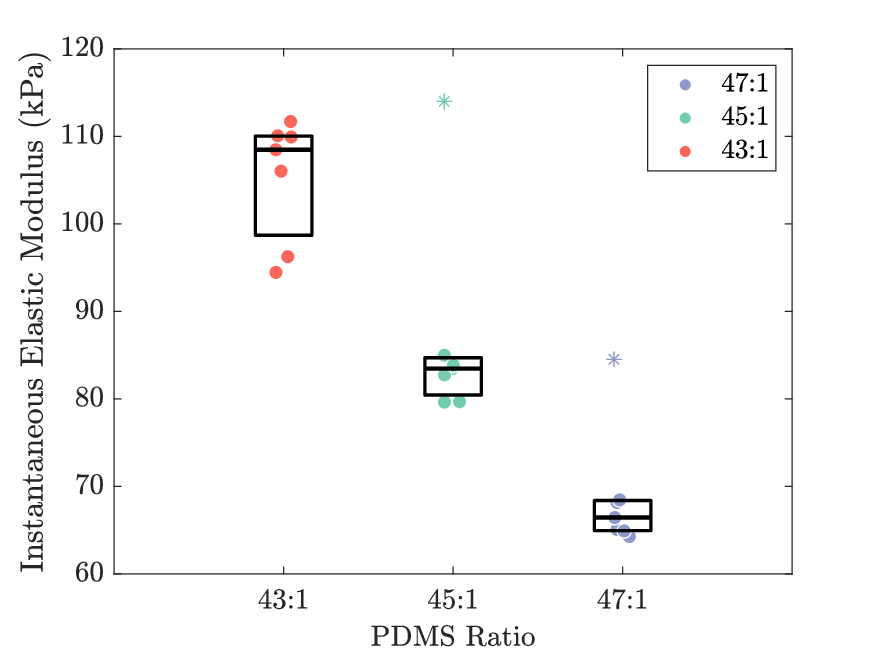}	
	\caption{Box plots illustrating the instantaneous elastic modulus, $E$, for the 43:1 (red), 45;1 (green), and 47:1 (blue) PDMS ratios.} 
	\label{fig_11}%
\end{figure}

\begin{figure}
	\centering 
	\includegraphics[width=0.5\textwidth, angle=0]{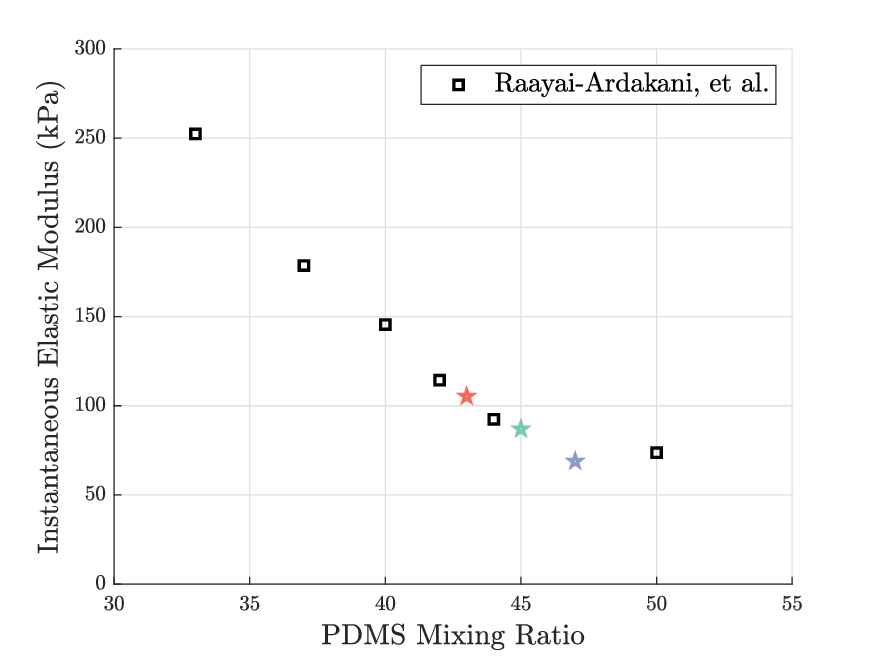}	
	\caption{Comparison of instantaneous elastic modulus, $E$, from validating this instrument against Raayai-Ardakani, et al\cite{raayai2019volume}.} 
	\label{fig_21}%
\end{figure}

In Fig. \ref{fig_11}, the fitted elastic moduli, $E$, demonstrate a decreasing trend as the base:curing agent ratio in PDMS increases, and more importantly, these values agree with existing literature, as seen from the comparison in Fig. \ref{fig_21}. The average elastic modulus values for each PDMS ratio are 105.28$\pm$7.02kPa for 43:1, 86.91$\pm$12.12kPa for 45:1, and 68.84$\pm$7.10kPa for 47:1, differentiating the mechanical properties of the samples, even at low sample sizes ($N$=7).

An argument could be made regarding whether the samples labeled 45-5 and 47-6 constitute outliers in comparison to the remainder of the dataset (Table \ref{table_4}). At first glance, these values might seem anomalous; however, revisiting the energy discussion from earlier and looking at both $U_c$ and $U_s$ (Fig. \ref{fig_8} and Table \ref{table_4}), we see their energies are in good agreement with one another with no visual outliers. This suggests that the observed discrepancy may not stem from instrumentation performance, but from extraneous factors. These include pre-test conditions, such as material lodged in the needle that adversely affects volumetric expansion, or through applying a constitutive model without extensive fine-tuning for each dataset.

In light of these considerations, the decision to retain these datasets in all calculations, rather than treat them as outliers was deliberate, even though their outlier treatment would greatly collapse the elastic moduli results: 43:1 equaling 105.28$\pm$7.02kPa, and 45:1 equaling 82.40$\pm$2.24kPa, and 47:1 equaling 66.23 $\pm$1.77kPa. This serves to underscore the inherent variability encountered in testing soft materials and further the application of constitutive models. Additionally, it highlights the potential utility of the energy metric as a cross-check for constitutive modeling or potentially even a stand-alone metric for material diagnostics.

\section{\label{sec:level6}Summary and Conclusions}
We have designed an innovative bench-top testing instrument for the VCCE method, specifically to advance the analysis of soft materials. This instrument stands as the first of its kind, transforming VCCE into a portable, user-friendly format suitable for space-constrained environments. Our development not only simplifies the VCCE testing process, but also ensures accessibility without compromising on the accuracy or depth of material analysis. By examining PDMS samples across a range of base:curing agent ratios, we validated instrument utility and simultaneously illustrate diagnostic utility of this instrument for broad adoption. Specifically, we have shown, this instrument may hold utility for biomedical applications in distinguishing between healthy and diseased tissue or understanding how material toughness varies between samples. 

The introduction of a bench-top VCCE instrument represents crucial advancement in the understanding of soft material mechanics. It facilitates deeper material property insights and promise new research avenues, aiming to improve both theoretical and practical approaches in soft material science. As we look ahead, the continued evolution of this technology and methodology will undoubtedly yield significant contributions to the field, enhancing our ability to explore and understand the complex world of soft materials.

\section*{\label{sec:level7}CRediT authorship contribution statement}
\textbf{Brendan M Unikewicz:} Writing – review \& editing, Writing – original
draft, Visualization, Validation, Software, Design, Methodology, Investigation,
Formal analysis, Data collection \& curation, Conceptualization
\textbf{Andre M Pincot:} Writing - review \& editing, Writing - draft methodology, figure captions, figure curation, Data collection \& curation
\textbf{Tal Cohen:} Writing – review \& editing, Supervision, Methodology, Funding acquisition, Conceptualization

\section*{\label{sec:level8}Declaration of Competing Interest}
The authors declare that they have no known competing financial
interests or personal relationships that could have appeared to influence
the work reported in this paper.
\section*{\label{sec:level9}Data Availability}
The data that support the findings of this study are openly available in our GitHub: 
github.com/cohen-mechanics-group/cots-benchtop-vcce

\section*{\label{sec:level10}Acknowledgements}
The authors express their gratitude to Chockalingam Senthilnathan, Hannah Varner and Katie Spaeth for their discussions on Volume-Controlled Cavity Expansion (VCCE) techniques. \newline 
\indent We acknowledge the partial support of our work through Office of Naval Research grant N000142312530, and the support from the National Science Foundation under award number CMMI1942016. A.M.P further acknowledges the Draper Scholar fellowship.

\bibliography{aipsamp}
\newpage
\appendix

\section{Boundary Effects in the neo-Hookean Model}

\renewcommand\thefigure{\thesection.\arabic{figure}}    
\setcounter{figure}{0}    
 \begin{figure}[h!]
	\centering 
	\includegraphics[width=0.5\textwidth, angle=0]{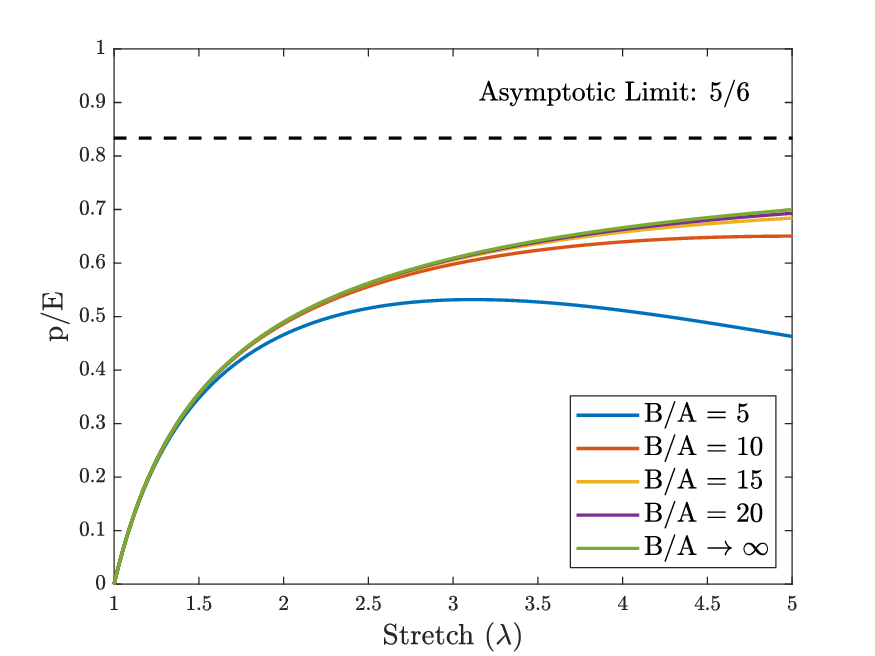}	
	\caption{The role of sample size highlighted with respect to effective initial defect size for the neo-Hookean hyperelastic model.} 
	\label{fig_2}%
\end{figure}
Raayai-Ardakani et al\cite{raayai2019volume}, showed the influence of boundary effects in the neo-Hookean model. Defining the circumferential stretch at the boundary as $\lambda_b$: 
\begin{equation}
    \lambda_b = \frac{b}{B} =\left[ \left(1 + \left(\lambda^3 - 1\right)\right) \left(\frac{A}{B}\right)^3 \right]^{1/3}.
    \label{eq_2}
\end{equation}
\indent The ratio, $B/A$, captures the relationship between the size of the boundary and the initial defect size. This ratio's significance is underscored in our investigation of boundary effects on material stretch during assumed cavitation events. To this end, we employed multiple $B/A$ ratios (as seen in Fig. \ref{fig_2}, aiming to elucidate their impact on the anticipated stretch values within various cavitation geometries:
\begin{equation}
    \frac{p}{E} = \frac{1}{6} \left( \lambda_b^{-4} + 4\lambda_b^{-1} - \lambda^{-4} - 4\lambda^{-1} \right).
    \label{eq_2A}
\end{equation}
\newpage
\section{Adverse Effects of Improper Syringe Selection}
\begin{figure}[h]
	\centering 
	\includegraphics[width=0.5\textwidth, angle=0]{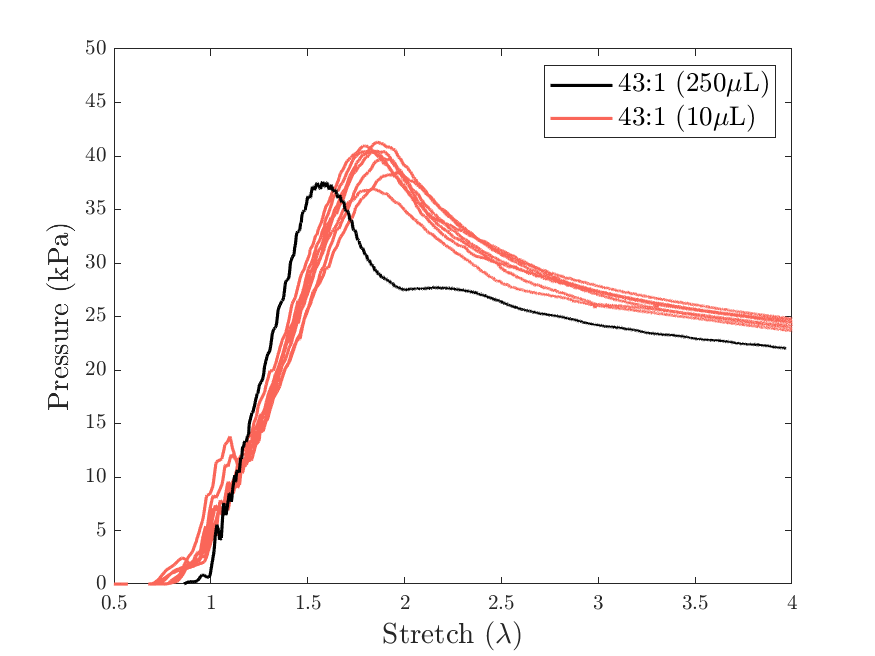}	
	\caption{The impact of varying the syringe size from 10$\mu$L (red) to 250$\mu$L (black) within VCCE protocols and how undesirable frequencies emerge and impact the measurement of soft materials.} 
	\label{fig_4}%
\end{figure}

To highlight the impact of increasing syringe diameter, a comparison between 250$\mu$L and 10$\mu$L syringes was performed. This comparison evaluated a 43:1 PDMS sample at $Q$=300nL/s. In Fig. \ref{fig_4}, the pressure-stretch response reveals a prominent, slowly varying signal of significant magnitude at low stretches, which dampens as the measurement progresses, suggesting a perceived increase in material stiffness due to large $\delta{V}$. This area is critical for the application of constitutive models, especially the derived result from the neo-Hookean model (Eq. \ref{eq_1}), hence the 10$\mu$L syringe was selected.

\newpage\section{Cumulative Energy Analysis}

\begin{figure}[h]
	\centering 
	\includegraphics[width=0.5\textwidth, angle=0]{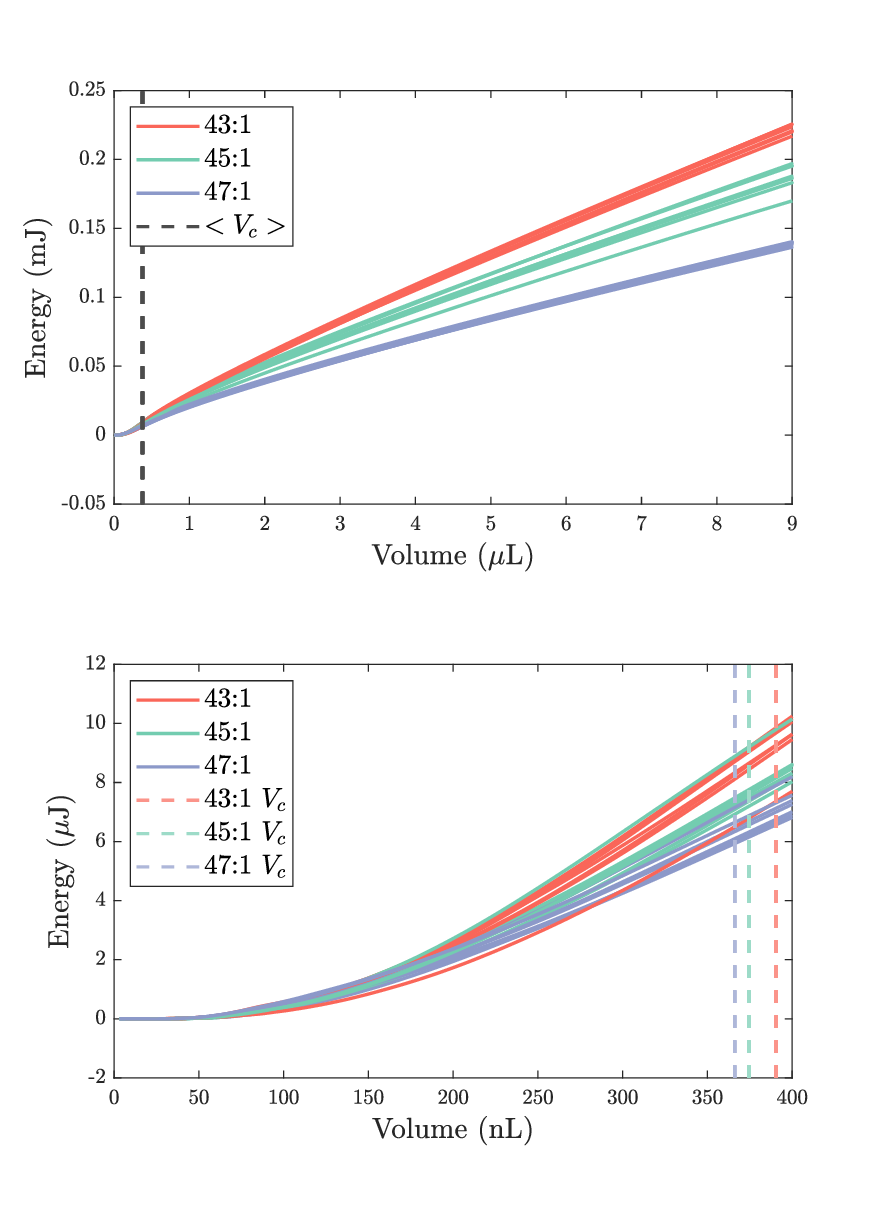}	
	\caption{Cumulative energy within the system evaluated through the nonlinear response of the material, $U_s$ (top), and available energy within the system evaluated up until fracture, $U_c$ (bottom), with average volumes of fracture for each PDMS ratio indicated.} 
	\label{fig_7}%
\end{figure}

Displaying the cumulative integration yielding $U_s$ and $U_c$ in Fig. \ref{fig_7}, with top plot depicting $U_s$, and bottom plot depicting $U_c$, we observe the energies of each sample converging as volume increases. Particularly, in the top figure, we observe that the average volume at which all samples fractured, $<V_c>$, represents only a small fraction of the total dataset. Further analysis of this minor portion reveals significant overlap in energies, with groupings beginning to form right at the average values of $V_c$ each material fractured.

\newpage\section{Needle Insertion Depths}
\vspace{-5cm}
The insertion procedure, broken down into distances the needle has traveled relative to the sample's surface is broken down in Table \ref{table_5}.
\vspace{-9cm}
\begin{table}[h]
\caption{Insertion Procedure Data -- Distance Traveled}
\centering 
\begin{ruledtabular}
\begin{tabular}{l c c c c} 
 Sample & $z_p$ (mm) \footnote{Distance traveled from surface until needle penetrates sample} & $z_{f}$ (mm) \footnote{Distance traveled from $z_p$ where needle is inserted deeper into sample} & $z_{r}$ (mm) \footnote{Distance traveled from $z_p + z_f$ where needle is raised towards surface} & $z_{t}$ (mm) \footnote{Total distance traveled from surface prior to performing VCCE test} \\ \hline
 43-1 & -12.0 & -5.0 & +5.0 & -12.0 \\
 43-2 & -15.0 & -5.0 & +4.0 & -16.0 \\
 43-3 & -15.0 & -5.0 & +5.0 & -15.0 \\
 43-4 & -13.0 & -5.0 & +5.0 & -13.0 \\
 43-5 & -12.0 & -5.0 & +4.0 & -13.0 \\
 43-6 & -15.0 & -5.0 & +4.0 & -16.0 \\
 43-7 & -15.0 & -5.0 & +4.0 & -16.0 \\ \hline
 45-1 & -15.0 & -5.0 & +5.0 & -15.0 \\
 45-2 & -15.0 & -5.0 & +4.0 & -16.0 \\
 45-3 & -16.0 & -5.0 & +4.0 & -17.0 \\
 45-4 & -15.0 & -5.0 & +5.0 & -15.0 \\
 45-5 & -17.0 & -5.0 & +4.0 & -18.0 \\
 45-6 & -15.0 & -5.0 & +4.0 & -16.0 \\
 45-7 & -15.0 & -5.0 & +4.0 & -16.0 \\ \hline
 47-1 & -17.0 & -5.0 & +5.0 & -17.0 \\
 47-2 & -17.0 & -5.0 & +4.0 & -18.0 \\
 47-3 & -16.0 & -5.0 & +5.0 & -16.0 \\
 47-4 & -15.0 & -5.0 & +5.0 & -15.0 \\
 47-5 & -16.0 & -5.0 & +4.0 & -17.0 \\
 47-6 & -15.0 & -5.0 & +4.0 & -16.0 \\
 47-7 & -17.0 & -5.0 & +4.0 & -18.0 \\
\end{tabular}
\end{ruledtabular}
\label{table_5}
\end{table}

\end{document}